\documentclass[onecollarge,natbib]{svjour2}
\bibpunct{[}{]}{;}{n}{}{,} 
\smartqed  
\usepackage{graphicx}
\usepackage{epsfig}
\usepackage{amsfonts}
\usepackage{amssymb}
\usepackage{amsmath}
\usepackage{latexsym}
\usepackage{braket}
\usepackage{bm}
\journalname{Few-Body Systems}
\newcommand{\beq}{\begin{eqnarray}}
\newcommand{\eeq}{\end{eqnarray}}
\newcommand{\be}{\begin{equation}}
\newcommand{\ee}{\end{equation}}
\def\la{\mathrel{\mathpalette\fun <}}

\def\fun#1#2{\lower3.6pt\vbox{\baselineskip0pt\lineskip.9pt
\ialign{$\mathsurround=0pt#1\hfil ##\hfil$\crcr#2\crcr\sim\crcr}}}

\newcommand{{\SD}}{\rm SD}

\newcommand{\vep}{\bm p}

\begin{document}

\title{The Regge trajectories and leptonic widths of the vector $s\bar s$ mesons}

\author{A.M.~Badalian \and B.L.G. Bakker}
\institute{A.M.~Badalian
\at
Institute of Theoretical and Experimental
Physics, Moscow, Russia,
\email{badalian@itep.ru}
\and
B.L.G. Bakker
\at
 Faculty of Science, Vrije Universiteit,
 Amsterdam, The Netherlands,
 \email{b.l.g.bakker@vu.nl}
 }

\date{Received: date / Accepted: date}

\maketitle

\begin{abstract}
The spectrum of the  $s\bar s$ mesons is studied performing a phenomenological analysis of the Regge trajectories
defined for the excitation energies. For the $\phi(3\,^3S_1)$ state the mass $M(\phi(3S))=2100(20)$~MeV and the
leptonic width $\Gamma_{ee}(\phi(3S))=0.27(2)$~keV are obtained , while the mass of  the $2\,^3D_1$ state,
$M(\phi(2\,^3D_3))=2180(5)$\,MeV, appears to be in agreement with the mass of the  $\phi(2170)$ resonance, and its
leptonic width, $\Gamma_{ee}(2\,^3D_1)=0.20\pm 0.10$\,keV,
has a large theoretical uncertainty, depending on the parameters of the flattened confining potential.
\end{abstract}

\maketitle

\section{Introduction}
Recently BES III observed a new resonant structure in the $J/\psi\rightarrow \phi\eta^{\prime}\eta$ decays in the
$\phi\eta^{\prime}$ invariant mass distribution, denoted as $X(2000)$ \cite{ref.1}. The quantum numbers of $X(2000)$
are not fixed yet and two possibilities are presented. First, assuming the state $J^{P}=1^-$, the mass $M(X)=2002.1\pm 27.5 \pm
15.0$~MeV and the width $\Gamma=129\pm 17\pm 7$\,MeV with the significance $5.3\sigma$ were obtained, while
assuming the state $J^{P}=1^{+}$, the larger mass $M(X)= 2062.8 \pm 13.1 \pm 4.2$~MeV of the structure with the
width $\Gamma=177\pm 36\pm 20$ ~MeV and the significance $4.9\sigma$ were determined.
This new structure was already analyzed in several theoretical studies, where in the conventional $s\bar s$ picture the resonance $X(2000)$
is considered as a candidate of the vector $\phi(3\,^3S_1)$ state \cite{ref.2} or, in Ref.~\cite{ref.3}, it was interpreted as
the second excitation of the axial-vector $h_1(1380)$ meson. A different conception of $X(2000)$ was suggested in Refs.~\cite{ref.4,ref.5}, where
$X(2000)$ is assumed to be a candidate of the $ss\bar s\bar s$ tetra-quark with $J^{PC}=1^{+-}$.

The $X(2000)$ together with $\phi(2170)$ represent a special interest for the theory, being the highest excitations in the
$s\bar s$ system observed up to now, although the properties and decays of the $s\bar s$ excitations were studied for
decades  \cite{ref.6,ref.7,ref.8,ref.9,ref.10,ref.11}  in different approaches: in the framework of relativistic potential models
(RPM) \cite{ref.7,ref.8}, the Regge trajectories (RTs) \cite{ref.9,ref.10}, and the QCD sum rules \cite{ref.11}. Comparison of the
predicted masses shows that the masses of  high $\phi(nS)$ excitations differ by $\sim (100-150)$~MeV, even if the
masses of the low states practically coincide. Such differences can be easily understood taking into account that the
properties of high excitations are very sensitive to the chosen values of the constituent quark masses and the
parameters of the quark-antiquark interaction. This statement can be illustrated by the masses of the $nS$ and $nD$ states, collected in
Ref.~\cite{ref.2} and given in Table 1, where MGI refers to the modified Godfrey-Isgur model with the screened confining
potential \cite{ref.2}.
\begin{table}
\caption{The masses of the $\phi(nS)$ and $\phi(n\,^3D_1)$ states (in MeV)\label{tab.1}}

\begin{tabular}{|c|c|c|c|c|}
\hline
  State      & GI\cite{ref.7}  & EFG \cite{ref.8}  & MGI \cite{ref.2}  &  experiment\cite{ref.12}\\
  \hline

 $\phi(1S)$   & 1016       &   1038         &  1030         &  1020  \\

 $\phi(2S)$     & 1687    & 1698         & 1687          &  1680(20)\\

 $\phi(3S)$     & 2200     &  2119       &  2149          &  2002(42) ? \\

 $\phi(4S)$   &   2622     & 2472           &   2498       &  abs.  \\

 $\phi(1D)$  &   1876       & 1845        & 1869           &  abs \\

$\phi(2D)$ &   2337        &  2258      &   2276            & 2188(10) ? \\

$\phi(3D)$  &   2725       &  2607       &   2593          &  abs. \\
\hline
\end{tabular}
\end{table}

From Table~\ref{tab.1} one can see that in all RPMs with constituent quark masses,  the mass $M(\phi(3S))$ appears to be by
$(120 - 200)$\,MeV larger than the mass of $X(2000)$ with $J^{P}=1^{-}$ in
experiment. Also in RPMs  the first excitation $\phi(2D)$ has a mass larger than the mass of $\phi(2170)$ by $\sim
100$\,MeV, although in the MGI model, where a screened confining potential (CP) is used,
the masses of the $\phi(3S)$ and $\phi(2D)$  states are about $50$\,MeV smaller than those in the GI model \cite{ref.7},
where the purely linear CP is used.
Note  that for $\phi(4S)$ and $\phi(3D)$ the GI model predicts values of the masses, which are  already by $\sim
120$~MeV larger, which means that within the same model the choice of the parameters of the confining potential at
large distances is crucially important to describe higher excitations. The situation is different, if an analysis of the spectrum is
performed with the help of Regge trajectories, defined for excitation energies, denoted as ERT \cite{ref.13}, where the
parameters of the ERT can be extracted from experiment and the predicted masses of high excitations appear to be
smaller  than those in RPMs.
In our paper we perform a phenomenological analysis of the $\phi(nS)$ and $\phi(nD)$ resonances,  using the
ERT, introduced by Afonin, Pusenkov \cite{ref.11}, as it was done in the analysis of the heavy-quarkonia spectra in
Ref.~\cite{ref.13}. We will also discuss how  the parameters of the ERT depend on the mass of the $s$-quark.

\section{The radial ERT of $\phi(nS)$ mesons}

In  heavy quarkonia, as well as in the $s\bar s$ system,  the  ERT are defined for the excitation energies
$E(nJ)$ \cite{ref.11},
\be
M(nJ) - 2m_s= E(nJ).  ~~\label{eq.1}
\ee
First, we consider the $\phi(n\,^3S_1)$ radial trajectory, which needs special consideration since its radial slope is larger
than that of the ERT for the states with $l\not= 0$ \cite{ref.13}; this effect is seen in the RT of light mesons \cite{ref.9,ref.10,ref.15,ref.16},
as well as in heavy quarkonia \cite{ref.13}. The reason why the radial slope is larger in the $S$-wave mesons
is explained by the stronger  gluon-exchange (GE) interaction in the states with $l=0$ than in those with $l\not= 0$ \cite{ref.14}.

The radial ERT of $\phi(nS)$ can be presented as,
\be
 E^2(nS)= a_{\rm S} + \beta_{\rm S}\, n_r. ~~ \label{eq.2}
\ee
where $n_r$ is the radial quantum number. The $s$-quark mass $m_s$, the intercept $a_{\rm S}$, and the slope
$b_{\rm S}$ can be extracted from experiment, if  there are enough  experimental data on the $\phi(nS)$ excitations,
measured with great accuracy. However, the existing experimental data do not allow to extract  $m_s$ at low scale and
here we take $m_s$ at low scale, using the relation for the running mass in pQCD \cite{ref.17,ref.18} and the conventional
value of  $m_s(q=2\,{\rm GeV}/c)=96(5)$\,MeV \cite{ref.12}, defined at the scale $q=2$\,GeV$/c$.
The sizes of high  $s\bar s$ mesons are large, $> 1.0 $ ~fm, and therefore their dynamics are determined by small characteristic momenta,
$q\la 1$\,GeV. The following mass relations,  $m_s(q=1\,{\rm GeV}/c)=1.27 m_s(q=2\,{\rm GeV}/c)=122$~MeV and
$m_s(q=0.5\,{\rm GeV}/c)=1.97\mu_s(q=2\,{\rm GeV}/c)=189$\,MeV were obtained in Ref.~\cite{ref.17}, i.e.,
$m_s\sim (120-180)$\,MeV at low scale. Just the value $m_s=180$~MeV was used in the analysis of the decay
constants of the strange mesons $D_s$ and $ B_s$ \cite{ref.18}.

Calculations with the use of the ERT show that in both cases, when $m_s(\sim 0.5\,{\rm GeV})=180$~MeV or
$m_s(\sim 1\,{\rm GeV})=125$~MeV are taken, the masses of the $\phi(nS)$ mesons differ  only within (10 - 20)\,MeV,
i.e., weakly depend on the chosen value of $m_s$, although the intercept and the radial slope of
the radial ERTs Eq.~(\ref{eq.2}) are different (see their values in Table 2). In Table 2  for comparison we give also the masses
$M(\phi(nS))$ from Ref.~\cite{ref.2}, where the MGI model with screened confining potential is used, giving the masses of the $3S$ and $4S$ excitations  by $\sim 50$~MeV larger than those in our analysis of the ERT.

\begin{table}
\caption{The masses of $\phi(nS)$ (in GeV) from the ERT and the parameters $a_{\rm S},~b_{\rm S}$ (in GeV$^2$)
for $m_s=0.125$\,GeV and $m_s=0.180$\,GeV and the masses
from Ref.~\cite{ref.2} and from experiment \cite{ref.12}\label{tab.2}}
\begin{tabular}{|c|c|c|c|c|}
\hline
         & $m_s=0.125$~GeV  & $m_s=0.180$~GeV & $m_s=0.387$~GeV \cite{ref.2} & experiment \\
\hline
   $a_s$       &  0.593         &   0.4356     &            &       \\

   $b_s$        &  1.465         &  1.30          &           &       \\

   $\phi(1S)$   &  1.020       & 1.020       &  1.030        &   1.020\\

   $\phi(2S)$  &  1.680      & 1.677        &  1.687         & 1.680(20)   \\

   $\phi(3S)$  &  2.121      &   2.102     &  2.149         & 2.002(42) ? \\

   $ \phi(4S)$  &2.477       &  2.442      &  2.498       & abs.  \\
   \hline
   \end{tabular}
   \end{table}
From Table~\ref{tab.2} one can see that the masses, defined by the  ERT with $m_s=125$\,MeV and $m_s=180$\,MeV,
coincide  within $\sim 10$\,MeV accuracy for $\phi(1S)$ and $\phi(2S)$ and differ only by $\sim 20$\,MeV for  the higher
$3S$ and $4S$ excitations, although their intercepts and the radial slopes are different. Notice that for a smaller
$m_s=125$\,MeV the values of the intercept and the radial slope are practically equal to those for light mesons
\cite{ref.10,ref.14,ref.16}, while for $m_s=180$~MeV they are closer to the values in heavy quarkonia \cite{ref.13}.

Notice that in our analysis, where $M(\phi(3S))=2102$\,MeV, as well as in RPMs (see Table~\ref{tab.1}), the mass of
$\phi(3\,^3S_1)$ is larger than the experimental mass of $X(2000)$ \cite{ref.1}.
However, one cannot exclude a large hadronic  shift-down of the $3\,^3S_1$ state due to the $P-$wave $\phi\phi$
threshold (with $M({\rm thresh.})=2039$~MeV) and then this state could be a candidate to be the  $X(2000)$ resonance,
as it is assumed in Ref.~\cite{ref.2}.

\section{The generalized ERT of the $s\bar s$ resonances}

The experimental masses of the ground states  $\phi(1S)$, $f_2^{\prime}(1525)$ and $\phi_3(1850)~(J=1,2,3)$ allow to define
the orbital parameter $b_J$ of the leading ERT with $J=l+1,~n_r=0$:
\be
E^2 (J,n_r=0) = a + b_{J}\, J.  ~~\label{eq.3}
\ee
First, we take $m_s=180$\,MeV and by definition of the ERT the mass differences can be written as
\be
M(\phi(1S)) -2m_s=\sqrt{a + b_J}=0.660\,{\rm GeV}; \sqrt{a + 2 b_J} - \sqrt{a+b_J}= 0.505\,{\rm GeV},
\label{eq.4}
\ee
giving the following intercept and  orbital slope of the leading ERT,

\be
 a = - 0.4644\,{\rm GeV}^2; ~~ b_J = 0.905\,{\rm GeV}^2,~~(m_s=0.180\,{\rm GeV},~ J=l+1).
 \label{eq.5}
\ee
From here the masses of the $s\bar s$ states with $J=l+1$ are following,
\begin{eqnarray}
 M(\phi(1S)) & = & 1.020\,{\rm GeV},\quad M(f_2^{\prime}(1P))=1.518\, {\rm GeV},
 \quad M(\phi_3(1D))= 1.859\,{\rm GeV}, \nonumber \\
 M(f_4(1F)) & = & 2.135\,{\rm GeV}, \quad M(\phi_5(1G))=2.374\,{\rm GeV},
 \label{eq.6}
\end{eqnarray}
where the masses of $\phi(1S)$, $f_2^{\prime}(1P)$, and $\phi_3(1D)$ are in precise agreement with the experimental masses of $\phi(1020)$, $f_2^{\prime}(1525)$, and $\phi_3(1850))$, respectively. At the same time, the  calculated mass $M(f_4(1P))=2131$\,MeV is  by $\sim 100$\,MeV larger than $M(f_4(2050))=2018(11)$\,MeV ($\Gamma=237\pm 18$~MeV) \cite{ref.12}. One can present arguments that the resonance $f_4(2050)$ has a large hadronic shift and therefore does not lie on a linear ERT (or on  a conventional RT). First, from the mass difference $\mu^2= M^2(f_4(2050)) - M^2(\phi_3(1850)=0.635$\,GeV$^2$ one obtains a very small orbital slope for the conventional RT, $\mu^2=0.635$\,GeV$^2$, and an even smaller slope $b_J= 0.517$\,GeV$^2$ for the ERT. Secondly, the mass difference between the $s\bar s$ ground state and corresponding ground state of a light meson is typically $\sim 200$\,MeV, e.g. $M(\phi(1020)) - M(\rho(775))=245$\,MeV, $M(f_2^{\prime}(1525) - M(a_2(1320))= 207$\,MeV, while the masses of $M(f_4(2050))$ and $M(a_4(2040)$ are almost equal (their mass difference is 23(18)\,MeV). This may occur if  $f_4(2050)$ has a large hadronic shift, possibly, due to the nearby  $\phi\phi$ threshold.

To describe the radial $s\bar s$ excitations with $l\not= 0$, we assume that $f_2(1950)$ is an $s\bar s$ state,  but not
assuming a priori that $\phi(2170)$ is the first excitation of $\phi_3$, since its quark structure is still discussed
\cite{ref.19,ref.20}. Then the radial slope of the ERT (for the states with $l\not= 0$) can be extracted from experiment
(the parameters $a, b_J$ are given in Eq.~(\ref{eq.5}), using the mass difference,
\be
\sqrt{a + 2\,b_J + b_n} - \sqrt{a+ 2\,b_J} = 1.944(12) - 1.525(5) = 0.419\pm 0.017\,{\rm GeV}.
\label{eq.7}
\ee
From here  the radial slope  $b_n= (1.15\pm 0.05)$\,GeV$^2$ is extracted and the generalized ERT,
\be
E^2(J,n_r) ({\rm in~GeV}^2) = - 0.4644  + 0.905\, J  + 1.15(5)\, n_r,~~ (J=l+1)
\label{eq.8}
\ee
gives the masses, presented in Table~\ref{tab.3}, where besides the masses of the $n\,^3D_3$, the masses of
the $n\,^3D_1$ states are given, which are by $\sim 15(5)$\,MeV smaller due to the fine-structure splitting.
\begin{table}
\caption{The masses (in MeV) of the $s\bar s$ states with $l\not= 0$ from ERT Eq.~(\ref{eq.7})\label{tab.3}}
\begin{center}
\begin{tabular}{|c|c|c|c|}
\hline
State     & $n_r=0$  &  $n_r=1$   & $n_r=2$   \\
\hline
$n\,^3 P_2$ &  1518    & 1938     & 2.268   \\

experiment & 1525(5) & 1944(12)  & abs.   \\

$n\,^3D_3$  &1858    &  2203    &  2492 \\

experiment &  1854(7)&  abs. &  abs. \\

$n\,^3D_1$   & 1835   &  2185    &  2475   \\

experiment  &  abs.  &   2188(10)? &  abs.   \\
\hline
\end{tabular}
\end{center}
\end{table}

In our calculations the mass $M(2\,^3D_1)$ coincides with that of $\phi(2070)$ and this fact indicates that $\phi(2070)$
can have large $s\bar s$ component, as it was assumed in Ref.~\cite{ref.20}.

\section {The leptonic widths}

To define the leptonic widths of the $\phi(1020), \phi(1680), \phi(1\,^3D_1), \phi(2\,^3D_1)$ their wave functions (w.f.s) are calculated, using the relativistic string Hamiltonian (RSH) \cite{ref.14,ref.15,ref.21} without fitting parameters, where the centroid mass $M_{\rm cog}(nl)$ is defined by the eigenvalues $M_0(nl)$ of the  equation,
\begin{equation}
 (2\sqrt{\vep^2+ m_s^2}  + V_0(r))\varphi_{nl}(r) = M_0 (nl) \varphi_{nl}(r).
 \label{eq.9}
\end{equation}
and by two negative corrections, the self-energy and the string corrections \cite{ref.14,ref.15,ref.22}. Here $m_s=180$\,MeV
and the potential $V_0(r)$ is taken as the  sum of the confining potential  and the gluon-exchange (GE) term,
$V_{\rm GE}=-\frac{4\alpha_{\rm V}(r)}{3 r}$, with the parameters from Ref.~\cite{ref.14}. Since the ground states
have relatively small sizes, their dynamics is defined by the linear CP, $V_{\rm c}(r)=\sigma r=0.18~r$\,GeV, while
for high excitations, which have large sizes, a flattened CP $V_{\rm f}(r)$,
\be
 V_{\rm f}(r)=\sigma_{\rm f}(r) r
 \label{eq.10}
\ee
has to be taken \cite{ref.14}. We assume that the self-energy and the string correction do not affect the w.f.s and the
radial w. f.s at the origin $R_{nS}(0)$ and $ R_{nD}(0)$ are defined by the solutions $\varphi_{nl}(r=0)$ of  Eq.~(\ref{eq.8}),
where for the $n\,^3D_1$ states the w.f.  $R_{nD}(0)$ is expressed via the derivative $R_{nD}^{\prime\prime}(0)$
\cite{ref.23},
\be
R_{nD}(0) = \frac{5}{2\sqrt{2}} \frac{R_{nD}^{\prime\prime}(0)}{\omega^2(nD)},~~\label{eq.11}
\ee
Here the flattened CP $V_{\rm f}(r)$ is chosen as in Ref.~\cite{ref.14} with the string tension,
$\sigma_{\rm f}(r)= \sigma (1 - \gamma f(r))$ and the function $f(r)$,
\be
 f(r) = \frac{\exp(\sqrt{\sigma}(r -R_0))}{ B + \exp(\sqrt{\sigma}(r-R_0))},
\label{eq.12}
\ee
defined by the following parameters,
\be
 \gamma =0.40, ~~B=20, ~R_0= 6.0~{\rm GeV}^{-1},~~ \sigma=0.182~{\rm GeV}^2.
\label{eq.13}
\ee
If the flattened CP+GE term  is used, then the sizes of the $nS$ and $nD$ excitations increase, see Table~\ref{tab.4},
where also  the w.f.s  $R_{nS}(0),~R_{nD}(0)$ and the kinetic energies $\omega(nL)$ of the $s-$quark, entering
$R_{nD}(0)$, are given. We have observed an interesting effect: if a flattened CP is used, then the kinetic energies
$\omega(nl)$ and the w.f.  $R_{nS}(0)~(n_r \geq 1)$ decrease, while the second derivatives $R_{nD}^{\prime\prime}(0)$
increases. Consequently, the leptonic width of the $\phi(nD)$ also increases. The leptonic width of a vector $s\bar s$
meson with the mass $M_{\rm V}(nl)~(l=0,2)$ (the charge squared $e_s^2=1/9,~\alpha=(137)^{-1}$) is given by the
expression \cite{ref.18},
\be
\Gamma_{ee}(\phi(nL))= \frac{4\,e_s^2 \alpha^2 |R_{nl}(0)|^2 ~\beta{\rm rel} ~\beta_{\rm QCD}}{ M_{\rm V}^2}, ~~\label{eq.14}
\ee
where the factor $\beta_{\rm rel}\cong (0.72-0.74)$ takes into account the relativistic effects \cite{ref.18}, and
$\beta_{\rm QCD}=1 -\frac{16\alpha_s(\mu)}{3\pi}=0.40$
is the QCD correction, where the strong coupling  $\alpha_s(\mu)=0.353$ at the scale $\mu\sim 1.4$~GeV is taken.

\begin{table}[!htb]
\caption{The r.m.s.~(in fm), the quark kinetic energy (in GeV), the w.f. at the origin $R_{nl}(0)$ (in GeV$^{3/2}$),
and the leptonic widths $\Gamma_{ee}(nl)$ (in keV) of the $s\bar s$ vector mesons ($m_s=0.180$~GeV)\label{tab.4}}
\begin{center}
\begin{tabular}{|c|c|c|c|c|}
\hline
     state &   r.m.s (LP) & $\omega(nl)$ & $R_{nl}(0)$ &  $\Gamma_{ee}$\\
\hline
     $1S$    & 0.65      &   0.477    &    0. 432  &   1.24 \\
     $2S$    &  1.15     &   0.596       & 0.410(10)  & 0.42(2)  \\
     $3S$    & 1.74       &  0.613    &  0.400       & 0.27(2)\\
     $4S$     &2.59       &  0.63      &   0.410    & 0.19 (2)\\
\hline
    $1D$     &  1.18       &0.603     &   0.25(8) & 0.13(9) \\
    $2D$     &  1.81    & 0.610      &   0.35(7) & 0.20 (10)\\
    $3D$     &  2.70     & 0.630     &   0.44(10) & 0.22(12) \\
\hline
\end{tabular}

\end{center}
\end{table}
In the GE potential  we use the vector coupling constant, which does not contain fitting parameters and takes
into account the asymptotic freedom behavior \cite{ref.24}, so that
the effective coupling of the ground state $\alpha_{\rm V}({\rm eff.})=0.39$ is relatively small, while
$\alpha_{\rm V}=0.54$ is larger for excited states with $n_r\geq 2$. Details can be found in Ref.~\cite{ref.14},
where the vector coupling $\alpha_{\rm v}(n_f=3)$ is shown to be defined via the QCD vector constant
$\Lambda_{\rm V}(n_f=3)=0.455$\,GeV, which corresponds to the QCD constant
$\Lambda_{\overline{MS}}(n_f=3)=330$\,MeV from Ref.~\cite{ref.25}.

\section{Conclusions}
The spectrum of the $s\bar s$ mesons was studied with the use of the ERT trajectories, defined for the excitation
energies, $E(nJ)=M(nJ) -2 m_s$ \cite{ref.11}. It is shown that the parameters of the ERT depend on the value of the
$s$-quark mass at a low scale. Two values, $m_s=125$\,MeV and $m_s=180$\,MeV, are considered.
In both cases the calculated  masses coincide  within $(10-20)$~MeV accuracy, although for $m_s=125$~MeV the slope
$b(nS)=1.465$\,GeV$^2$ and the intercept $a(nS)=0.593$\,GeV$^2$ are larger than those for  $m_s=180$~MeV,
and equal to the parameters of the $\rho(nS)$ RT. If $m_s=180$~MeV is taken, the values $b_n(nS)=1.30$\,GeV$^2$ and
$a(nS)=0.4356$\,GeV$^2$ are smaller and close to those in heavy quarkonia \cite{ref.13}. For $\phi(3S)$ the
leptonic width $\Gamma_{ee}=0.42(2)$\,keV is obtained.

With the use of the ERT the predicted masses of the high excitations appear to be smaller than those calculated in
potential models with a constituent $s$-quark mass. For $\phi(3S)$ the calculated mass $M(\phi(3S))=2100(20)$\,MeV is
larger than that of the $X(2000)$ resonance, recently observed by  BES III, but a large hadronic shift down of this
resonance is not excluded. For the states with $l\not= 0$ the generalized ERT, which includes  the orbital and radial excitations, has the orbital slope $b_J=0.905$\,GeV$^2$ and the radial slope $b_n=1.15(5)$\,GeV$^2$. This ERT gives
the mass $M(f_2(2P))= 1938$\,MeV in agreement with the mass of $f_2(1950)$ and $M(f_2(3P))=2268$\,MeV, while the
mass $M(2\,^3D_1)=2.180(5)$\,GeV agrees with the mass of the $\phi(2170)$ resonance, and therefore $\phi(2170)$
could be either the $2\,^3D_1$ state or contain a large $s\bar s$ component. The leptonic width of $\phi(2D)$, $
\Gamma_{ee}=0.20(10)$\,keV, has a large theoretical uncertainty, which occurs because of the strong
sensitivity of the radial w.f. at the origin $R_{2D}(0)$ to the parameters of  flattened (screened) potential.


\begin{thebibliography}{99}

\bibitem{ref.1}
M.~Ablikim et al. (BESIII Collab.), ``Observation and study of $J/\psi \rightarrow \phi\eta\eta^{\prime}$ at BES III",
arXiv:1901.00085 [hep-ex].

\bibitem{ref.2}
C.~Q.~Pang, ``The excited states of $\phi$ mesons", arXiv: 1902.02206(2019),  [hep-ph].

\bibitem{ref.3}
L.~M.Wang, S.~Q.~Luo, and X.~Liu , ``$X(2100)$ newly observed in $J/\psi\rightarrow \phi\eta\eta^{\prime}$ at BES III
as an isoscalar axial-vector meson", arXiv:1901.00636 (2019), [hep-ph].

\bibitem{ref.4}
E.~L.~Cui et al., ``QCD sum rule studies of the $ss\bar s\bar s$ tetraquark states with $J^{PC}=1^{--}$",
arXiv:1901.01724 (2019), [hep-ph].

\bibitem{ref.5}
Z.~G.~Wang, ``Light tetraquark state candidates", arXiv:1901.04815 (2019), [hep-ph].

\bibitem{ref.6}
T.~Barnes, N.~Black, and P.~R.~Page, ``Strong decays of Strange Quarkonia", Phys. Rev. {\bf D 68}, 054014 (2003);
nucl-th/0208072 (2002).

\bibitem{ref.7}
S.~Godfrey and N.~Isgur. ``Mesons in relativized quark model with chromodynamics", Phys. Rev. {\bf D 32}, 189 (1985).

\bibitem{ref.8}
D.~Ebert, R.~N.~Faustov, and V.~O.~Galkin, ``Mass spectra and Regge trajectories of light mesons in the relativistic
quark model", Phys. Rev. {\bf D 79}, 114029 (2009), arXiv:0903.5183 [hep-ph] (2009).

\bibitem{ref.9}
A.~V.~Anisovich, V.~V.~Anisovich, and A.~V.~Sarantsev, ``Systematics of $q\bar q$ states in the $(n,M^2)$ and
$(J,M^2)$ planes", Phys. Rev. {\bf D 62}, 051502 (2000), hep-ph/0003113;
V.~V.~Anisovich, ``Systematics of quark-antiquark states and scalar exotic mesons", Phys. Usp.{\bf 47}, 45 (2004),
hep-ph/0208123 (2002);
A.~V.~Anisovich et al., ``Light $2^{++}$ and $0^{++}$ mesons",  Phys. Rev. {\bf D 85}, 014001 (2012);
arXiv:1110.4333 (2011).

\bibitem{ref.10}
D.~V.Bugg,  ``Comments on `Systematics of radial and angular-momentum Regge trajectories of light non-strange
$q\bar q$ states' '', Phys. Rev. {\bf D 87}, 118501 (2013).

\bibitem{ref.11}
S.~S.~Afonin,``Towards understanding broad degeneracy of non-strange mesons",
 Int. J. Mod. Phys. {\bf A 22}, 1359 (2007); hep-ph/0701089 (2007);
 ``Properties of new unflavored mesons",  Phys. Rev. {\bf C 76}, 015202 (2007), arXiv:0707.0824;
 S.~S.~Afonin and I.~V.~Pusenkov, ``Universal description of radially excited heavy and light mesons",
 Phys. Rev. {\bf D 90}, 094020 (2014), arXiv:1411.2390 (2014) and references therein.

\bibitem{ref.12}
G.~Patignani et al. (PDG), Chin. Phys. {\bf C 40}, 100001 (2016).

\bibitem{ref.13}
A.~M.~Badalian and B.~L.~G.~Bakker, ``Radial and orbital Regge trajectories in heavy quarkonia",
arXiv:1902.09174 (2019) [hep-ph].

\bibitem{ref.14}
A.~M.~Badalian and B.~L.~G.Bakker, ``Dynamics of the quark-antiquark interaction and the universality of
Regge trajectories", arXiv:1901.10280 (2019) [hep-ph].

\bibitem{ref.15}
A.~M.~Badalian and B.~L.~G.~Bakker, ``The radial Regge trajectories and leptonic widths of the isovector mesons",
Phys. Rev. {\bf D 93},  074034 (2016), arXiv:1603.04725 (2016);
``Light meson excitations in the QCD string approach", Phys. Rev.  {\bf D 66}, 034025 (2002), hep-ph/0202246.

\bibitem{ref.16}
P.~Masjuan, E.~ R.~Arriola, and W.~Broniowsky, ``Radial and angular-momentum Regge trajectories:
a systematic approach", Phys. Rev. {\bf D 85}, 094006 (2012),
arXiv:1305.3493; ``Reply to `Comments on "Systematics of radial and angular momentum Regge trajectories of light non-
strange  $q\bar q$ states' ",
Phys. Rev. {\bf D 87},  118502 (2013).

\bibitem{ref.17}
A.~M.~Badalian and B.~L.~G.~Bakker, ``The running mass $m_s$ at low scale from the heavy-light decay constants",
JETP Lett. {\bf 86}, 634 (2008); arXiv:hep-ph/0702229 (2007).

\bibitem{ref.18}
A.~M.~Badalian, B.~L.~G.~Bakker, and Yu.~A.~Simonov, ``Decay constants of the heavy-light mesons in the field
correlator method", Phys. Rev.{\bf D 75}, 116001 (2007).

\bibitem{ref.19}
H.~W.~Ke and X.~Q.~Li, ``Study of the strong decays of $\phi(2170)$ and a grand expectation for the future
charm-tau factory", Phys. Rev. {\bf D 99}, 036014 (2019) and references therein, arXiv:1810.07912 (2018).

\bibitem{ref.20}
S.~Coito, G.~Rupp, and E. van Beveren,  ``Multichannel calculation of excited vector $\phi$ resonances and the
$\phi(2170)$'', Phys. Rev. {\bf D 80}, 094011 (2009), arXiv:0909.0051 (2009).

\bibitem{ref.21}
A.~Yu.~Dubin, A.~B.~Kaidalov, and Yu.~A.~Simonov, ``Dynamical regime of the QCD string with quarks",
Phys. Lett. {\bf B 323}, 41 (1994);
``The QCD string with quarks. 1. Spinless quarks", Phys. Atom. Nucl. {\bf 56}, 1745 (1993);
(Yad. Fiz. {\bf 56}, 213 (1993)).

\bibitem{ref.22}
Yu.~A.~Simonov, ``Nonperturbative corrections to the quark self-energy", Phys. Lett. {\bf 515}, 137 (2001);
A.~Di~Giacomo and Yu.~A.~Simonov, ``The quark-gluon mixed condensate calculated via field correlators", Phys. Lett. {\bf B 595}, 368 (2004).

\bibitem{ref.23}
V.~A.~Novikov et al., Phys. Rept. {\bf C 41}, 1 (1978);
A.~M.~Badalian and I.~V.Danilikin, ``Di-electron and two-photon widths in charmonium",
Phys. Atom. Nucl. {\bf 72}, 638 (2008), arXiv:0801.1614.

\bibitem{ref.24}
A.~M.~Badalian and B.~L.~G.~Bakker, ``The vector coupling $\alpha_{\rm V}(r)$ and the scales $r_0,r_1$
from the bottomonium spectrum",
Phys. Atom. Nucl. {\bf 77}, 767(2014), Yad. Fiz. {\bf 77}, 810 (2014),arXiv:1303.2815 (2013).

\bibitem{ref.25}
S.~ Bethke, ``World summary of $\alpha_s$ 2012", Nucl. Phys. Proc. Suppl. {\bf 234}, 220 (2013), arXiv:1210.0325.

\end{thebibliography}
\end{document}